# Ultra-low lattice thermal conductivity in $Cs_2BiAgX_6$ ($X$=Cl, Br): Potential thermoelectric materials

Enamul Haque and M. Anwar Hossain*

Department of Physics, Mawlana Bhashani Science and TechnologyUniversity, Santosh,Tangail-1902, Bangladesh

**Abstract**

We have explored electronic and thermoelectric properties of bismuth-based double-perovskite halides $Cs_2BiAgX_6$ by using first principles calculations. The calculated indirect bandgaps 2.85 eV and 1.99 eV for $Cs_2BiAgCl_6$ and $Cs_2BiAgBr_6$, respectively well agree with the measured value (2.77 eV of $Cs_2BiAgCl_6$ and 2.19 eV of $Cs_2BiAgBr_6$). We have calculated the relaxation time and lattice thermal conductivity by using relaxation time approximation (RTA) within supercell approach. The lattice thermal conductivities for both compounds are remarkably low and the obtained values at 300K for $Cs_2BiAgCl_6$ and $Cs_2BiAgBr_6$ are 0.078 and 0.065 $Wm^{-1}K^{-1}$, respectively. Such quite low lattice thermal conductivity arises due to low phonon group velocity in the large weighted phase space and large phonon scattering. The large Seebeck coefficient obtained for both halides at 400K. We have obtained the maximum power factors at 700K and the corresponding thermoelectric figure of merit for $Cs_2BiAgCl_6$ and $Cs_2BiAgBr_6$ are 0.775 and 0.774, respectively. The calculated results reveal that both halides are potential thermoelectric materials.

1. Introduction

The lead free double-perovskite halides have attracted a great interest due to their application in environmentally-friendly perovskite solar cells. Lead-halides efficiently absorb light in the visible range [1], have long diffusion length [2], as well as the photons emitted in the radiative recombination process are recyclable [2]. These lead-based halides have the lack of structural stability and when the cells (making with these halides) are exposed to light, the temperature and humidity are increased [3–7]. Due to this limitation, the lead (Pb) has been successfully replaced by Bi and Ag [8–10]. This substitution leads to form a double-perovskite halide. The experimental synthesis of two compounds $Cs_2BiAgCl_6$ and $Cs_2BiAgBr_6$ was successful and these two compounds have a highly tunable bandgap in the visible range [8–10]. The optical measurement and band structure calculation reveal that both halides are indirect bandgap semiconductors [8–10]. Many studies have been reported on the measurement (also calculation) of bandgap as well as optical properties of both compounds to reveal the suitability of application in the photovoltaic solar cells. However, the possibility of these halides as the thermoelectric materials is not studied yet. The efficiency of thermoelectric materials is determined by dimensionless figure of merit ZT defined as [11] $ZT = \frac{S^2\sigma}{\kappa}T$, where $S$ is the thermopower, $\sigma$ is the electrical conductivity and $\kappa$ is the thermal conductivity consisting of lattice and electronic part. A material with ZT ~1.0 is considered as good thermoelectric material. Such high ZT value may be obtained when the power factor $S^2\sigma$ is high and thermal conductivity is low. In general, semiconductors can exhibit such trend. $Cs_2BiAgCl_6$ and $Cs_2BiAgBr_6$ are face-centered cubic crystals with experimental lattice parameters 10.777 Å and 11.264 Å for $Cs_2BiAgCl_6$ and $Cs_2BiAgBr_6$, respectively [9, 12]. For $Cs_2BiAgBr_6$, no phase transition was observed until 703K [8].

In this paper, we have presented first-principles calculations of electronic properties, transport properties, lattice thermal conductivity as well as relaxation time of $Cs_2BiAgX_6$ halides. The lattice thermal conductivity has been found to be remarkably low in both compounds and the total thermal conductivity become small. The calculated results reveal that both compounds are promising for thermoelectric applications.

## 2. Computational details

Elastic and electronic properties were studied by using the full potential linearized augmented plane wave (LAPW) implemented in WIEN2k [13]. For good convergence, a plane wave cutoff of kinetic energy $RK_{max}$ =8.0 and $(15 \times 15 \times 15)$ $k$-point in Brillouin zone integration were selected. The muffin tin radii 2.5 for Cs, Bi, Ag, 2.15 for Cl, and 2.45 for Br, were used. The modified TB-mBJ [14] functional was used in the electronic structure and transport properties calculations. We have also performed these calculations including spin-orbit coupling. The convergence criteria of energy and charge were set to $10^{-4}Ry$ and 0.001e, respectively. The transport properties were calculated by BoltzTraP code [15]. For this, we used $(43 \times 43 \times 43)$ $k$-point in WIEN2k to generate the required input files and the chemical potential was set to the zero temperature Fermi energy. By solving the semi-classical Boltzmann transport equation, we can easily calculate the transport coefficients. The transport coefficients are defined in the Boltzmann transport theory [16–18] as

$$\sigma_{\alpha\beta}(T,\mu) = \frac{1}{V}\int \Sigma_{\alpha\beta}(\varepsilon)\left[-\frac{\partial f_\mu(T,\varepsilon)}{\partial \varepsilon}\right]d\varepsilon \qquad (1)$$

$$S_{\alpha\beta}(T,\mu) = \frac{1}{eTV\sigma_{\alpha\beta}}\int \Sigma_{\alpha\beta}(\varepsilon)(\varepsilon-\mu)\left[-\frac{\partial f_\mu(T,\varepsilon)}{\partial \varepsilon}\right]d\varepsilon \qquad (2)$$

$$K^e_{\alpha\beta}(T,\mu) = \frac{1}{e^2TV}\int \Sigma_{\alpha\beta}(\varepsilon)(\varepsilon-\mu)^2\left[-\frac{\partial f_\mu(T,\varepsilon)}{\partial \varepsilon}\right]d\varepsilon \qquad (3)$$

where V is the volume of a unit cell, $\alpha$ and $\beta$ represent Cartesian indices, $\mu$ is the chemical potential and $f_\mu$ Fermi-Dirac distribution function. The energy projected conductivity tensors can be calculated by using the following equation

$$\Sigma_{\alpha\beta}(\varepsilon) = \frac{e^2}{N}\sum_{i,k}\tau_{i,k}v_\alpha(i,k)v_\beta(i,k)\frac{\delta(\varepsilon-\varepsilon_{i,k})}{d\varepsilon} \qquad (4)$$

where N is the number of k-points for BZ integration, $i$ is the index of band, $v$ and $\tau$ represent the electrons group velocity and relaxation time, respectively. In BoltzTraP program, the constant relaxation time approximation (CRTA) is used.

To calculate lattice thermal conductivity, we used supercell approach creating total 412 displacements and $16\times 16\times 16$ mesh for BZ integration by using Phono3py code [19]. The required forces were calculated by the plane wave pseudopotential method in Quantum Espresso [20]. For this, 204 eV kinetic energy cutoff for wavefunctions and 816 eV for charge density were used. The convergence threshold for selfconsistency was set to $10^{-14}$ to obtain well converged basis set. The criterion of force convergence was set to $10^{-4}eV/Å$. In the force calculation, ultrasoft pseudopotential (generated by PS library 1.00) and PBE functional were utilized. This method of calculation for lattice thermal conductivity and other phonon related properties has been successfully used for many materials and found to be reliable [21–24]. The lattice part of the thermal conductivity can be calculated by solving linearized phonon Boltzmann equation (LBTE) [25], utilizing the single-mode relaxation-time (SMRT) method and expressed as [19]

$$\kappa = \frac{1}{NV}\sum_\lambda C_\lambda \boldsymbol{v}_\lambda \otimes \boldsymbol{v}_\lambda \tau^{SMRT} \tag{5}$$

where V is the volume of a unit cell, $\boldsymbol{v}$ is the group velocity and $\tau$ is the SMRT for the phonon mode λ. The mode dependent phonon heat capacity $C_\lambda$ can be calculated by the following equation

$$C_\lambda = k_B \left(\frac{\hbar\omega_\lambda}{k_BT}\right)^2 \frac{\exp(\hbar\omega_\lambda/k_BT)}{[\exp(\hbar\omega_\lambda/k_BT)-1]^2} \tag{6}$$

where $k_B$ is the Boltzmann constant. The group velocity can be expressed as

$$v_\alpha(\lambda) = \frac{1}{\omega_\lambda}\sum_{\kappa\kappa'\beta\gamma} W_\beta(\kappa,\lambda)\frac{\partial D_{\beta\gamma}(\kappa\kappa',\boldsymbol{q})}{\partial q_\alpha} W_\gamma(\kappa',\lambda) \tag{7}$$

where W is the polarization and D is the dynamical matrix. In this method, the relaxation time is considered to be equal to the phonon lifetime and given by [19]

$$\tau_\lambda^{SMRT} = \frac{1}{2\Gamma(\omega_\lambda)} \tag{8}$$

where $\Gamma(\omega_\lambda)$ is the phonon linewidth. The Grüneisen parameter can be calculated by the equation [19, 26]

$$\gamma(\boldsymbol{q}\nu) = -\frac{V}{2[\omega(\boldsymbol{q}\nu)]^2}\langle e(\boldsymbol{q}\nu)\left|\frac{\partial D(\boldsymbol{q})}{\partial V}\right|e(\boldsymbol{q}\nu)\rangle \tag{9}$$

where D($\boldsymbol{q}$) is the dynamical matrix, and e($\boldsymbol{q}\nu$) is the phonon eigen vector at the wave vector $\boldsymbol{q}$.

## 3. Results and discussions

Fig. 1 shows the crystal structures of $Cs_2BiAgX_6$ (X=Cl, Br). $Cs_2BiAgCl_6$ and $Cs_2BiAgBr_6$ are face-centered cubic crystal with space group $Fm\bar{3}m$ (#225) [9]. The occupied Wyckoff positions by Cs, Bi, Ag and $X$ atoms are 8c, 4a, 4b and 24e (x=0.2513 for Cl and 0.2504 for Br), respectively [9]. In all calculations, we used experimental lattice parameters 10.777 Å and 11.264 Å for $Cs_2BiAgCl_6$ and $Cs_2BiAgBr_6$, respectively [12].

### 3.1. *Mechanical properties*

The study of mechanical properties of materials are very important for device application. The different mechanical properties such as anisotropy, hardness, ductility, can be derived from elastic constant. The mechanical strength of a material is described by the elastic moduli and Poisson's ratio. The calculation details of elastic moduli can be found in the standard articles [27–29]. The melting temperature of a material can be predicted by the expression [30], $T_m = [553K + (5.91K/GPa)c_{11}] \pm 300K$. Our calculated elastic constants, moduli of elasticity, Poisson's ratio, and predicated melting temperature of $Cs_2BiAgX_6$ are presented in Table-1. The necessary and sufficient conditions of mechanical stability for a cubic crystal system are given as [31]

$$C_{11} - C_{12} > 0 \; ; \; C_{11} + 2C_{12} > 0 \; ; \; C_{44} > 0 \tag{10}$$

Table-1: Elastic constant and moduli of elasticity in GPa and Poisson's ratio of $Cs_2BiAgX_6$.

| Compound | $c_{11}$ | $c_{12}$ | $c_{44}$ | B | G | E | ν | B/G | $T_m \pm 300$ |
|---|---|---|---|---|---|---|---|---|---|
| $Cs_2BiAgCl_6$ | 69.01 | 16.51 | 10.63 | 34.01 | 15.41 | 40.17 | 0.303 | 2.20 | 960 |
| $Cs_2BiAgBr_6$ | 67.6 | 8.76 | 6.59 | 28.37 | 12.63 | 33.01 | 0.306 | 2.24 | 952 |

It is clear that both compounds are elastically stable. Both materials are ductile as found from both Cauchy pressure ($c_{11}$-$c_{44}$) [32] and Pugh ratio (B/G) [33]. The shear elastic anisotropy can be calculated by using the expression [34], $A = \frac{2c_{44}}{c_{11}-c_{12}}$. The calculated values of $A$ for $Cs_2BiAgCl_6$ and $Cs_2BiAgBr_6$ are 0.4 and 0.22, respectively and indicate the elastically anisotropic nature. The Vickers hardness can be theoretically found by using Pugh ratio (G/B) in the equation [35] $H_V = 2\left(\left(\frac{G}{B}\right)^2 G\right)^{0.585} - 3$. Our calculated Vickers hardness $H_V$ (0.94 GPa for $Cs_2BiAgCl_6$ and 0.43 for $Cs_2BiAgBr_6$) indicates the materials to be relatively low hard. The Debye temperature is related with lattice vibration and hence thermal stability. The Debye temperature can be calculated by the following equation [36]

$$\theta_D = \frac{h}{k_B}\left(\frac{3N}{4\pi V}\right)^{1/3}\left[\frac{1}{3}\left(\frac{2}{v_t^3} + \frac{1}{v_l^3}\right)\right]^{-\frac{1}{3}} \qquad (11)$$

where, $v_l$ and $v_t$ are given by $v_l = \left(\frac{3B+4G}{3\rho}\right)^{1/2}$, $v_t = \left(\frac{G}{\rho}\right)^{1/2}$. The calculated Debye temperatures are 202 and 162K for $Cs_2BiAgCl_6$ and $Cs_2BiAgBr_6$, respectively. The low Debye temperature of both compounds implies that the lattice thermal conductivity should be small.

### 3.2. Electronic properties

The double perovskite, $Cs_2BiAgX_6$ halides have the indirect bandgap as shown in Fig. 2. The inclusion of spin-orbit interaction splits the first conduction band and lifts down the conduction band to 2.47 eV at Γ-point. This creates a heavy but narrow electron band with an energy separation 1.23 eV from next two higher conduction bands. Our calculated band structure of both compounds are consistent with others calculation [8,9,12]. The dispersive conduction bands with a typical bandwidth ~4.0 eV in TB-mBJ (but 4.66 eV in TB-mBJ+SOC), arise from Bi-6s states due to the delocalized nature of these states in $Cs_2BiAgCl_6$. However, the bandwidth is ~3.15 eV in TB-mBJ (~3.95 eV in TB-mBJ+SOC) for $Cs_2BiAgBr_6$. The maximum peak of the valence band at X-point mainly comes from Ag-4d and Cl-3p states. The conduction bands due to Bi-6s states have higher energy (green line) than these for Ag-5s states. This leads to another the conduction state at L-point. The delocalized nature of Bi-6s states leads to the even parity of maximum conduction band at L point. However, bandgap energy state at the L-point has the lowest energy and thus also leads to odd parity. Therefore, transitions allowed by parity conservation should be observed. The valence bands are flatter than conduction bands. The flat nature of these bands comes from the strong hybridization of Cl/Br-3p/5p, and Ag-4d orbitals as these clearly can be seen from the projected density of states (PDOS) shown in Fig. 3. The Ag-4d, Bi-6p, and Cl/Br-3p/4p states have the dominant contributions to the density of states. The Bi-5d states have a very small contribution and are not shown in the Fig. 3. It is found that spin-orbit coupling makes conduction band flatter. A comparison of experimentally measured bandgap and theoretically calculated bandgap by using different functionals are presented in Table-2. Our calculated bandgaps for both compounds agree fairly with experimental results. However, the experimentally measured bandgap of both compounds is also different due to different structural relaxations used in the measurement. The calculation of bandgap including spin-orbit coupling reduces the indirect bandgap, from 2.85 to 2.51 eV for $Cs_2BiAgCl_6$ and 1.99 to 1.82 eV for $Cs_2BiAgBr_6$. Therefore, it is essential to include spin-orbit coupling effect in the band structure calculation for the correct explanation of energy band edges of these two halides. It is clear that GW [37] and TB-mBJ+SOC calculations provide more accurate bandgap comparatively than others functional calculation for $Cs_2BiAgX_6$.

Table-2: Comparison of experimental and calculated bandgap (in eV) using different functionals.

|  | EXP-I | EXP-II | EXP-III | PBE0 | HSE+SOC | GW | TB-mBJ | TB-mBJ+SOC |
|---|---|---|---|---|---|---|---|---|
| $Cs_2BiAgCl_6$ | 2.2 | 2.77 | 2.2 | 2.7 | 2.62 | 2.4 | 2.85 | 2.51 |
| $Cs_2BiAgBr_6$ | - | 2.19 | 1.9 | 2.3 | 2.06 | 1.8 | 1.99 | 1.82 |
| Ref. | [10] | [9] | [12] | [10] | [9] | [12] | This | This |

The calculated band structures of both compounds indicate that the indirect bandgap arises from Ag-4d and Cl-3p/Br-4p states which can be clearly understood from the calculated projected density of states (PDOS) as shown in Fig. 4. The Ag-4d and Cl-3p/Br-4p orbitals have the dominant contribution to the density of states for both compounds. The highest peak (around -1.9 eV) in the DOS comes from the sigma bonding combinations of Cs-5p, Bi-6s,and Bi-6p orbitals. The second peak (around -2.4 eV) arises from the strong hybridization of Ag-4d, Bi-5d (are not shown) and Cl-3p orbitals. The details descriptions of bonding, antibonding and spin-orbit coupling splitting of Bi-p states from the density of sates have been described by Eric T. McClure et al. [9].

### 3.3. *Thermoelectric transport properties*

The phonon group velocity with frequency is shown in Fig. 4. We see that the phonon group velocity in both compounds is quite low as expected (since Debye temperature is very low). The group velocity of $Cs_2BiAgCl_6$ is larger than that of $Cs_2BiAgBr_6$. The large weighted phase space of both compounds may be responsible for such low phonon group velocity. To find the amount of anharmonicity in both double-perovskite halides, the phonon Grüneisen parameter with frequency is illustrated in Fig. 5. The large value of Grüneisen parameter indicates the high anharmonicity in the crystal and hence the large phonon scattering. Our calculated phonon Grüneisen parameters of both compounds are very large and indicate the large phonon scattering in the both halides. Such intrinsic large phonon scattering leads to large weighted phase space and hence low phonon group velocity. Thus, the lattice thermal conductivity of both compounds is expected quite low. The calculated relaxation time at different temperature isillustrated in the Fig. 6. The relaxation times are 0.24 and 0.23ps at 300 K for $Cs_2BiAgCl_6$ and $Cs_2BiAgBr_6$, respectively. Note that relaxation time decreases sharply upto 100 K and after then decreases very slowly. Such large relaxation time arises narrow electron linewidths of the conduction bands (see Fig. 2). The calculated lattice thermal conductivity is remarkably small. At ambient temperature, the lattice thermal conductivities are 0.078 $Wm^{-1} K^{-1}$ and 0.065 $Wm^{-1} K^{-1}$ for $Cs_2BiAgCl_6$ and $Cs_2BiAgBr_6$, respectively. The quite low phonon group velocity in the large weighted phase space and large phonon scattering give rise to such low lattice thermal conductivity. The lattice thermal conductivity decrease with temperature as most of the heat is transferred by lowest frequencies phonons. The thermoelectric transport properties of $Cs_2BiAgX_6$ are presented in Fig. 7 and Fig. 8. The Seebeck coefficient, *S* for both compounds increases up to 400K due to its higher intrinsic activation energy than Fermi energy and after this activation level, the Fermi level is shifted by the temperature to the middle of the forbidden gap. After 400K, thus, the Seebeck coefficient decreases with temperature. The calculated Seebeck coefficient of $Cs_2BiAgCl_6$ is slightly increased up to 500K due to the inclusion of spin-orbit coupling as shown in the Fig. 7(a). This is expected since the conduction band becomes flattered due to the inclusion of spin-orbit interaction. The maximum Seebeck coefficient of $Cs_2BiAgCl_6$ and $Cs_2BiAgBr_6$ obtained at 400K are 240 and 241 $\mu V/K$, respectively. This high Seebeck coefficient arises from the large bandgap with flat conduction band near the Fermi level. The positive Seebeck coefficient ($S$) implies that both compounds are p-type materials. It is found that there is no spin-orbit effect on the Seebeck coefficient of $Cs_2BiAgBr_6$. The electrical conductivity ($\sigma/\tau$) increases with temperature (see Fig. 7b and Fig. 8b) indicating the semiconducting nature of $Cs_2BiAgX_6$. The calculated electrical conductivities ($\sigma$) for $Cs_2BiAgCl_6$ and $Cs_2BiAgBr_6$ at 300K are $4.83 \times 10^5$ $Sm^{-1}$ and $4.54 \times 10^5$ $Sm^{-1}$, respectively. The narrow heavy electron conduction band give rise such high electrical conductivity. The inclusion of spin-orbit coupling (SOC) effect slightly reduces the electrical conductivity of $Cs_2BiAgCl_6$ although bandgap reduces by SOC. The SOC effect reduces the density of states near the Fermi level and hence decreases electron-phonon scattering. The electronic part of the thermal conductivity increases slowly than electrical conductivity as shown in Figs. 7(c) and 8(c). This is because of the increase of phonon frequency with temperature. The total thermal conductivity for both compounds increases with temperature are presented in Figs. 7(d) and 8(d). The room temperature thermal conductivities for $Cs_2BiAgCl_6$ and $Cs_2BiAgBr_6$ are 11.02 and 10.49 W/mK, respectively. The variation of calculated power factors, $S^2\sigma$ with temperature are illustrated in Fig. 9. The power factor of both compounds is high and the values at 700K are 36.29 and 35.27 $mWm^{-1}K^{-2}$ for $Cs_2BiAgCl_6$ and $Cs_2BiAgBr_6$, respectively. The temperature dependence of thermoelectric figure of merits are shown in Fig. 10. We have calculated *ZT* values from 200 to 700K temperature range and the maximum value ~0.8 obtained at 700K for both compounds. We observed that spin-orbit coupling effect has not significant effect on the thermoelectric performance calculation. The high values of thermoelectric performance arises from the high electrical conductivity and low thermal conductivity of both compounds. This high thermoelectric figure of merits for $Cs_2BiAgX_6$ indicating that both compounds are suitable for thermoelectric energy conversion application.

## 4. Conclusions

In summary, the double perovskite $Cs_2BiAgX_6$ halides are found to be elastically stable, ductile and relatively low hard material. The calculated indirect bandgaps are 2.85 and 1.99 eV for $Cs_2BiAgCl_6$ and $Cs_2BiAgBr_6$, respectively and these values are in well agreement with the experimental values. We have calculated the relaxation time and lattice thermal conductivity by using RTA within supercell approach. The lattice thermal conductivity is extremely low due to quite low phonon group velocity in the large weighted phase space and large phonon scattering, and these are 0.078 and 0.065 $Wm^{-1}K^{-1}$ at 300 K for $Cs_2BiAgCl_6$ and $Cs_2BiAgBr_6$, respectively. The total thermal conductivities are 10.49, and 10.48 $Wm^{-1}$ $K^{-1}$ at 300 K for $Cs_2BiAgBr_6$ and $Cs_2BiAgCl_6$, respectively. The maximum thermopower obtained at 400K for $Cs_2BiAgCl_6$ and $Cs_2BiAgBr_6$ were 240 and 241 $\mu V/K$, respectively. The maximum power factors obtained at 700K were 36.3 and 35.2 $mWm^{-1}K^{-2}$ for $Cs_2BiAgCl_6$ and $Cs_2BiAgBr_6$, respectively. The thermoelectric figure of merits for $Cs_2BiAgCl_6$ and of $Cs_2BiAgBr_6$ compounds exhibit the similar trend with temperature and in both cases the maximum ZT value ~0.8 obtained at 700 K, implies that these halides are potential candidate for thermoelectric device applications. We hope our calculated thermoelecric properties of $Cs_2BiAgX_6$ (*X*=Cl, Br) will inspire experimentalist for further improvement by nanostructuring and grain refinement to reduce the thermal conductivity without affecting the electrical conductivity.

**Figures**

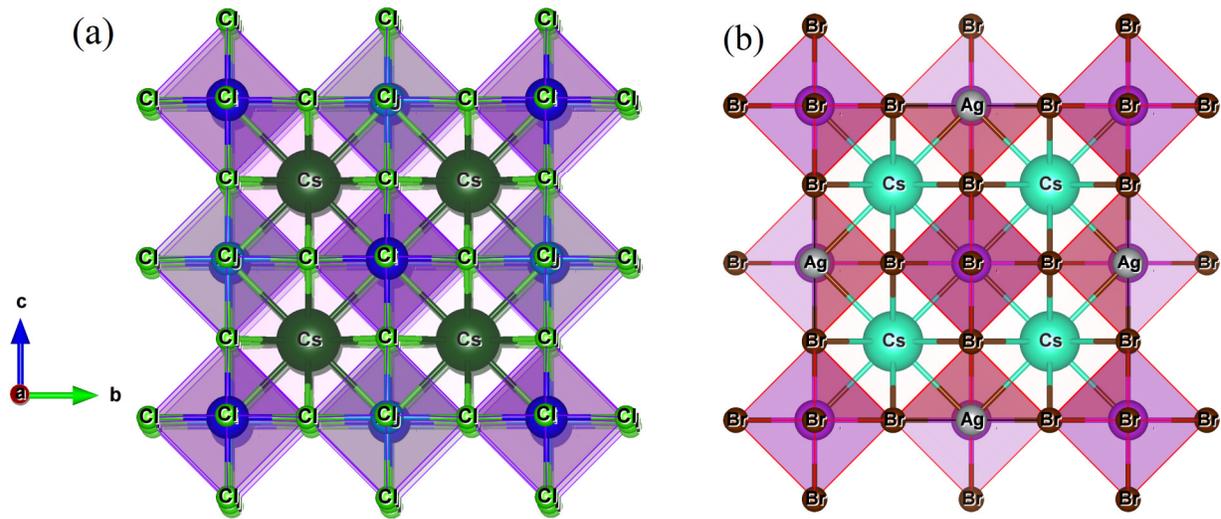

Fig. 1: Crystal structure of: (a) Cs$_2$BiAgCl$_6$, (b) Cs$_2$BiAgBr$_6$.

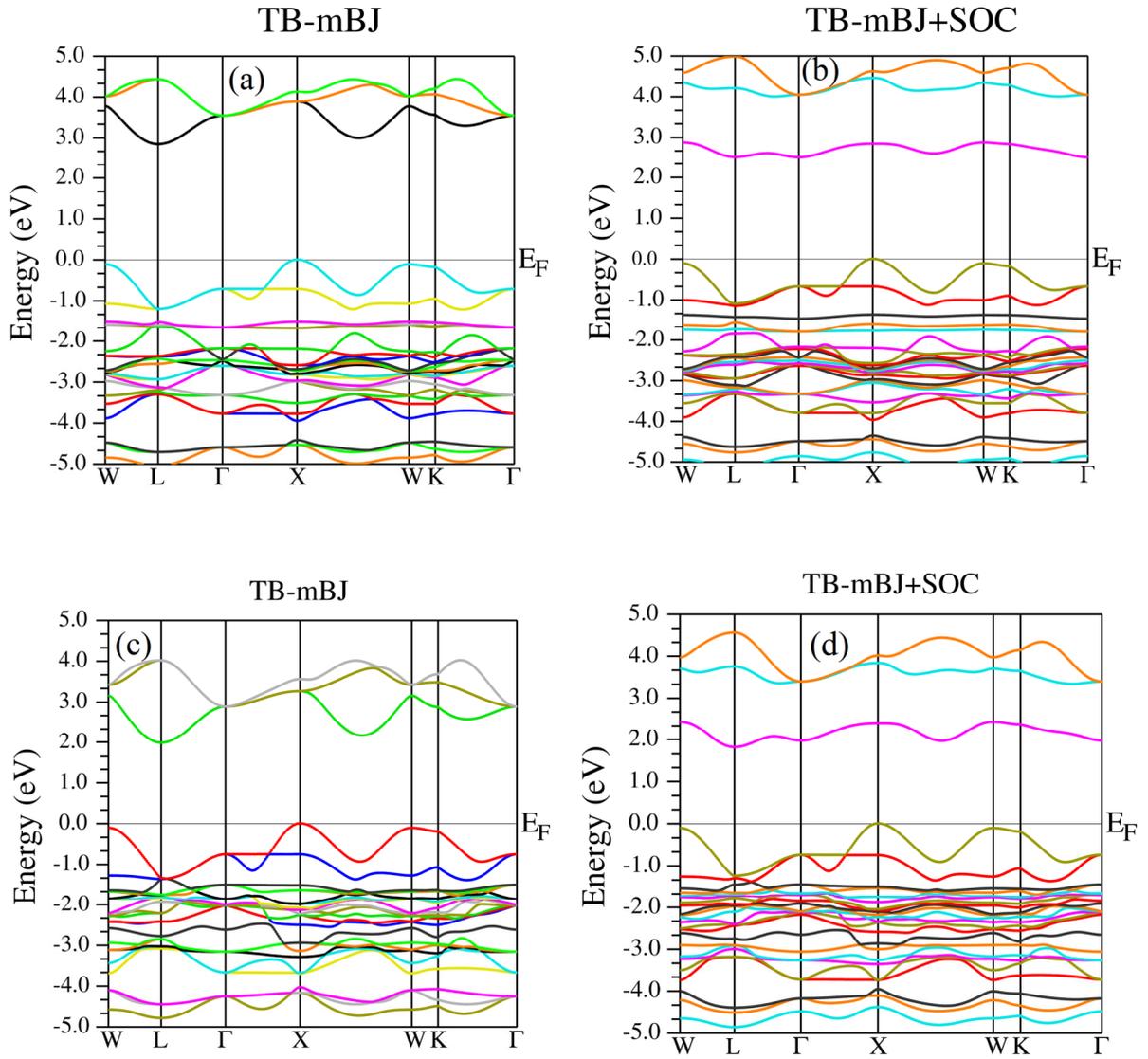

Fig. 2: The band structure of $Cs_2BiAgCl_6$, (a) without and (b) with spin-orbit coupling, and $Cs_2BiAgBr_6$, (c) without and (d) with spin-orbit coupling.

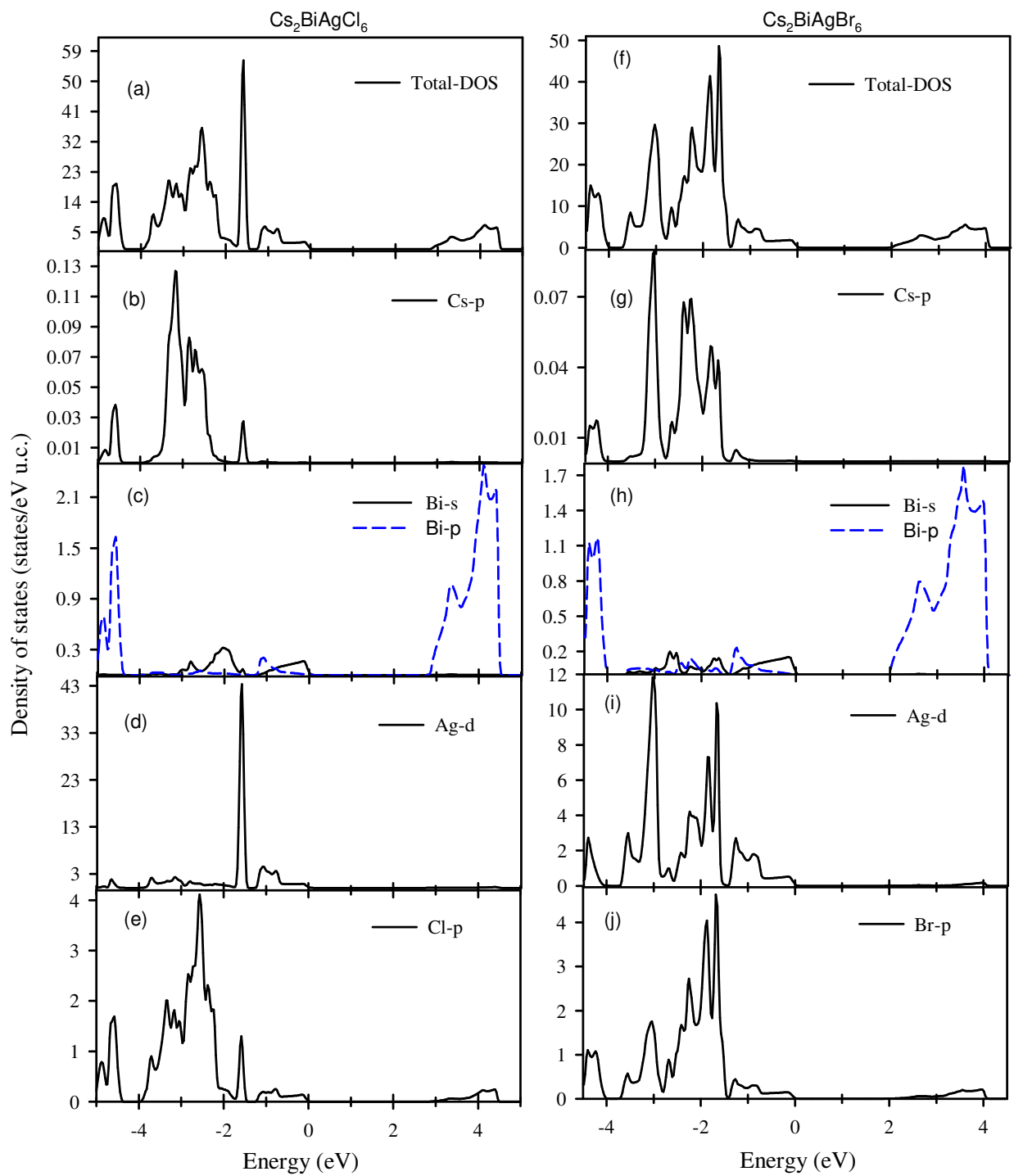

Fig. 3: Total and projected density of states of Cs$_2$BiAg$X_6$ without spin-orbit coupling.

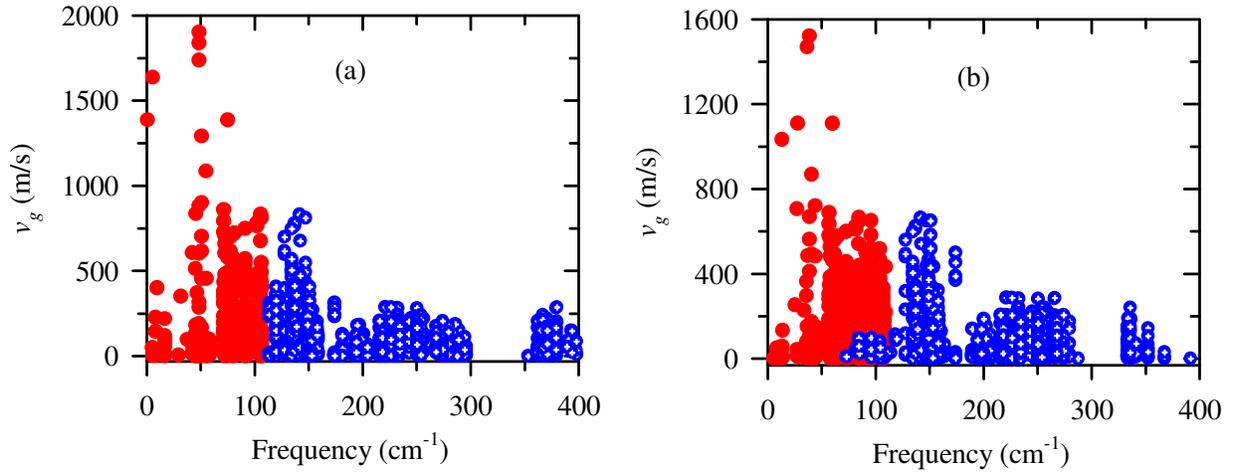

Fig. 4: Variation of calculated phonon group velocity with frequency: (a) $Cs_2BiAgCl_6$ and (b) $Cs_2BiAgBr_6$. The red circle indicates the acoustic mode, and blue circle (with cross) indicates optical mode.

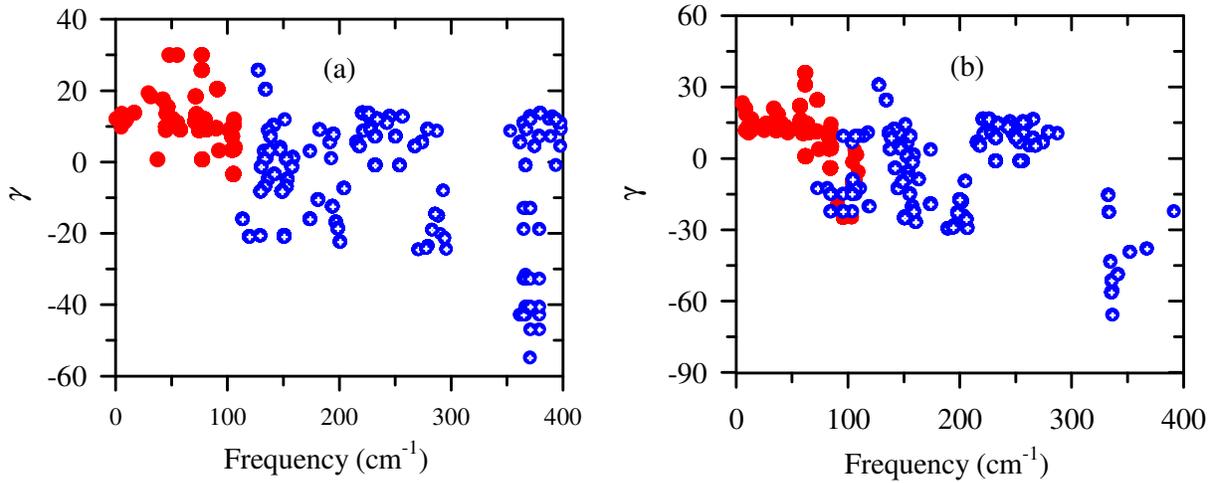

Fig. 5. Variation of phonon Grüneisen parameter with frequency: (a) $Cs_2BiAgCl_6$ and (b) $Cs_2BiAgBr_6$. The red circle indicates the acoustic mode, and blue circle (with cross) indicates optical mode.

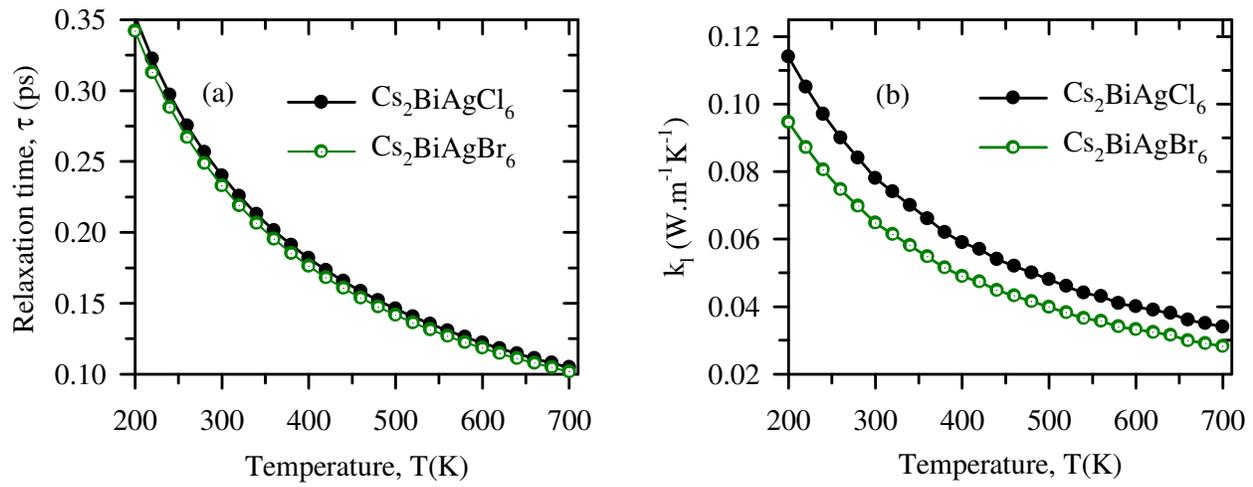

Fig. 6: (a) Relaxation time and (b) lattice thermal conductivity of $Cs_2BiAgX_6$.

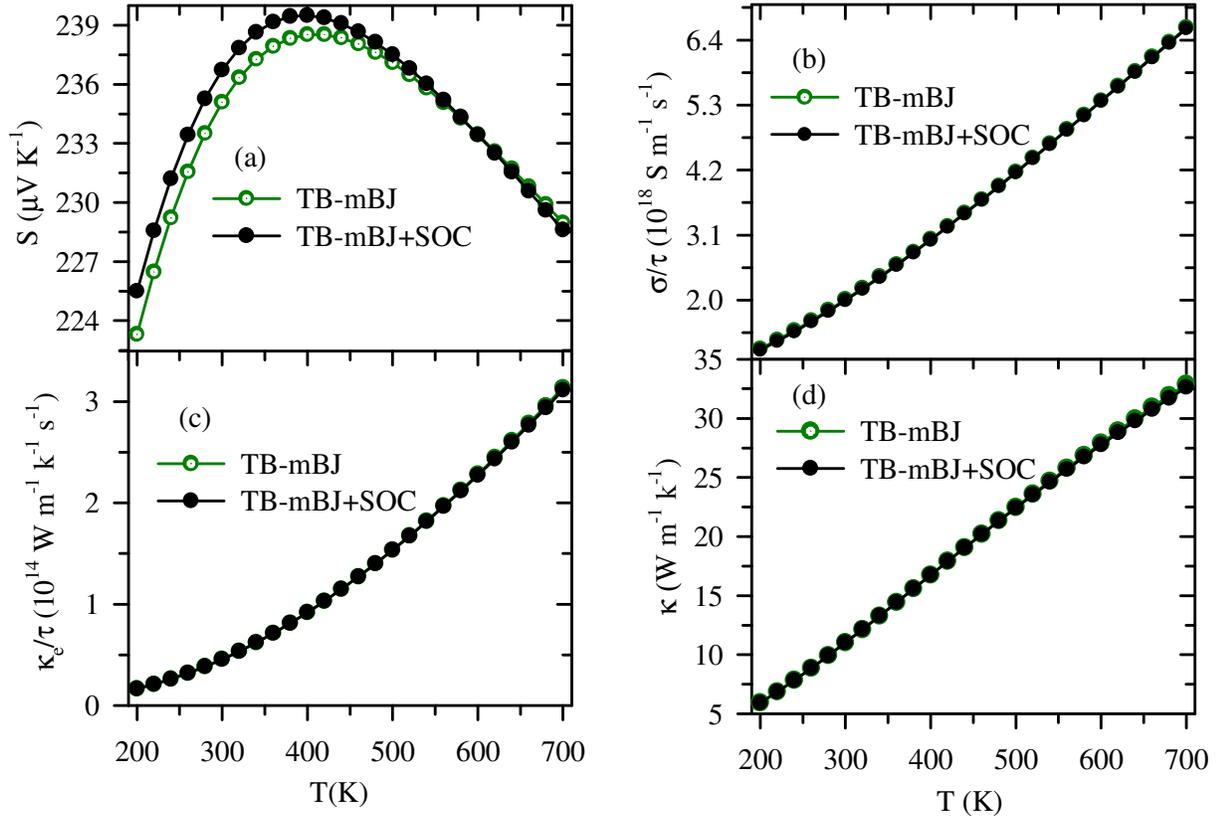

Fig. 7: Thermoelectric transport properties of $Cs_2BiAgCl_6$: (a) Seebeck coefficient (S), (b) electrical conductivity ($\sigma/\tau$), (d) electronic part of the thermal conductivity ($\kappa_e/\tau$), and (e) total thermal conductivity ($\kappa$).

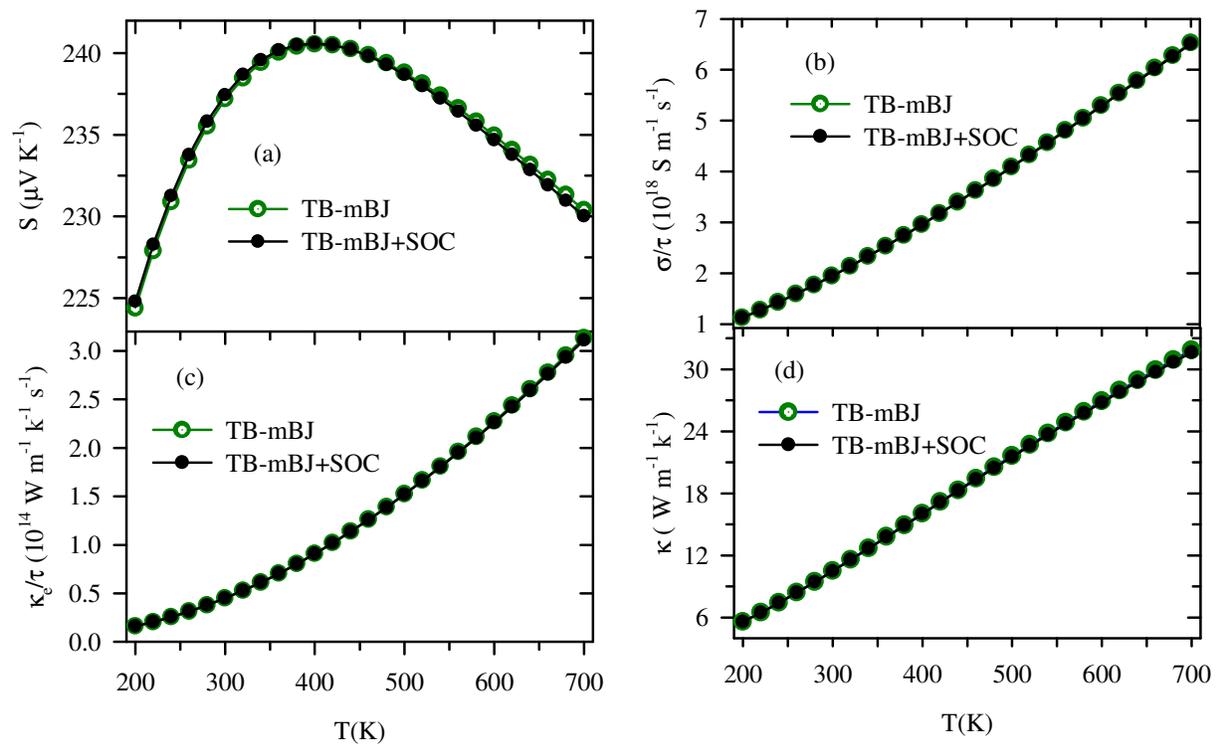

Fig. 8: Thermoelectric properties of $Cs_2BiAgBr_6$: (a) Seebeck coefficient ($S$), (b) electrical conductivity ($\sigma/\tau$), (d) electronic part of the thermal conductivity ($\kappa_e/\tau$), and (e) total thermal conductivity ($\kappa$).

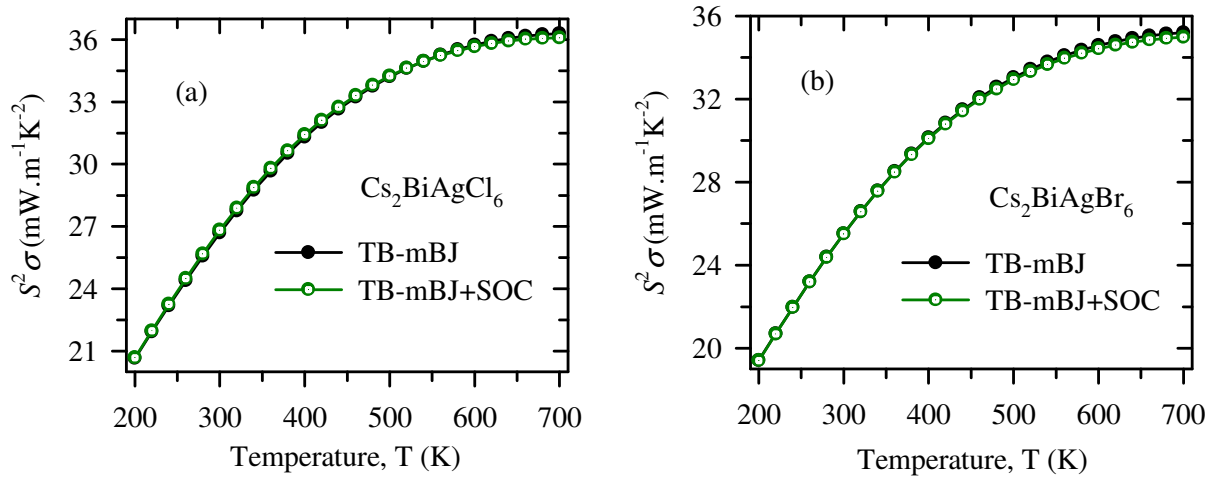

Fig. 9: Variation of calculated power factors $S^2\sigma$ of: (a) $Cs_2BiAgCl_6$ (b) $Cs_2BiAgBr_6$

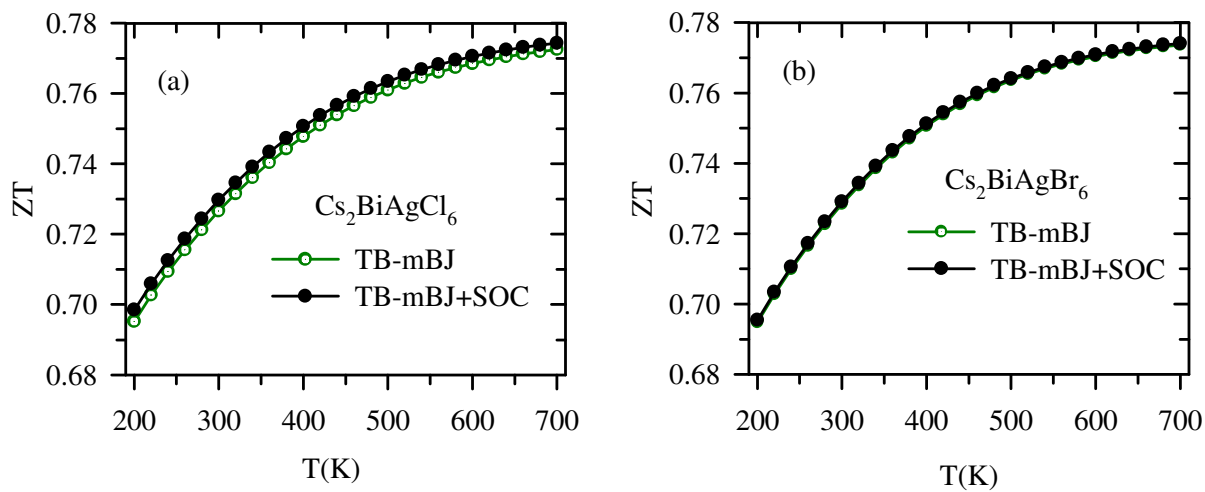

Fig. 10: Thermoelectric figure of merit of: (a) $Cs_2BiAgCl_6$ and (b) $Cs_2BiAgBr_6$.